\documentclass[twocolumn,amsmath,amssymb,showpacs]{revtex4}

\usepackage{enumitem}
\usepackage[dvips]{graphicx}
\usepackage[dvipdfm]{hyperref}
\usepackage{footmisc}
\usepackage{subfigure}
\usepackage{amssymb,amsmath}
\usepackage{appendix}
\usepackage{color}
\usepackage{helvet}
\DeclareFontFamily{U}{mathx}{\hyphenchar\font45}
\DeclareFontShape{U}{mathx}{m}{n}{
      <5> <6> <7> <8> <9> <10>
      <10.95> <12> <14.4> <17.28> <20.74> <24.88>
      mathx10
      }{}
\DeclareSymbolFont{mathx}{U}{mathx}{m}{n}
\DeclareFontSubstitution{U}{mathx}{m}{n}
\DeclareMathAccent{\widecheck}{0}{mathx}{"71}
\DeclareMathAccent{\wideparen}{0}{mathx}{"75}

\def\be{\begin{equation}}
\def\ee{\end{equation}}
\def\bea{\begin{eqnarray}}
\def\eea{\end{eqnarray}}

\begin{document}

\title{Gain, directionality and noise in microwave SQUID amplifiers: Input-output approach}
\author{Archana Kamal,$^{1}$ John Clarke,$^{2}$ and Michel Devoret$^{1}$}
\affiliation{
$^{1}$Departments of Applied Physics, Yale University, New Haven, CT 06520, USA\\
$^{2}$Department of Physics, University of California, Berkeley, CA 94720, USA}
\date{\today}

\begin{abstract}
We present a new theoretical framework to analyze microwave amplifiers based on the dc SQUID. Our analysis applies input-output theory generalized for Josephson junction devices biased in the running state. Using this approach we express the high frequency dynamics of the SQUID as a scattering between the participating modes. This enables us to elucidate the inherently nonreciprocal nature of gain as a function of bias current and input frequency. This method can, in principle, accommodate an arbitrary number of Josephson harmonics generated in the running state of the junction. We report detailed calculations taking into account the first few harmonics that provide simple semi-quantitative results showing a degradation of gain, directionality and noise of the device as a function of increasing signal frequency. We also discuss the fundamental limits on device performance and applications of this formalism to real devices. 
\end{abstract}

\pacs{85.25.Cp, 74.50.+r, 85.25.Dq, 85.25.-j} 
\maketitle

\section{Introduction}
\label{sec_one}
%
For almost half a century, the dc SQUID (Superconducting QUantum Interference Device) has enabled a broad range of devices, including magnetometers, gradiometers, voltmeters, susceptometers and amplifiers \cite{SQUIDvol1,SQUIDvol2}. Most of these devices are used at relatively low frequencies, and all have the common feature of offering extremely low noise \cite{Tesche}. The fact that the dc SQUID is potentially a quantum-limited amplifier in the microwave regime was recognized long ago \cite{koch:380}, but not exploited in practice until the Axion Dark Matter eXperiment (ADMX) provided a powerful motivation \cite{RevModPhys.75.777}. This need led to the development of the Microstrip SQUID Amplifier (MSA) in which the input coil deposited on (but insulated from) the washer of a SQUID acts as a resonant microstrip \cite{muck:2885}. Such amplifiers have achieved a noise temperature within a factor of two of the standard quantum limit \cite{Caves,muck:967,kinion:202503}. More recently, new designs have appeared intended to extend the frequency of operation to frequencies as high as 10 GHz, aimed at the readout of superconducting qubits \cite{PhysRevLett.107.053602} and the detection of micromechanical motion \cite{Motion}. These include incorporation of a gradiometric SQUID at the end of a quarter wave resonator \cite{spietz:082506} and the direct injection of the microwave signal from a quarter wave resonator into one arm of the SQUID ring \cite{ribeill:103901,hover:063503}.   
\par
Besides having desirable properties such as high gain, wide bandwidth and near quantum-limited operation, microwave SQUID amplifiers (MWSAs)---unlike conventional Josephson parametric amplifiers \cite{castellanos-beltran:083509, yamamoto:042510, BergealJPC}---also offer an intrinsic separation of input and output channels of the signal that makes them unique among  amplifiers based on Josephson tunnel junctions. This property makes them especially well suited as a preamplifier in the measurement chain for superconducting devices by eliminating the need for channel separation devices, such as circulators and isolators, between the sample under test and the first amplification stage. Although microwave SQUID amplifiers have been successfully used experimentally, questions pertaining to their nonlinear dynamics and ultimate sensitivity as amplifiers have continued to remain challenging problems. Previous theories include quantum Langevin simulations \cite{koch:380,Zorin} and treatment of the SQUID as an interacting quantum point contact \cite{Clerk}. The ultimate exploitation of the amplifier, however, requires a deeper understanding of its behavior at the Josephson frequency and its harmonics. Besides being valuable for practical considerations, such understanding may help discern the cause of intrinsically nonreciprocal operation of the MWSA that has hitherto remained an open question. This concern is especially relevant to applications such as qubit readout where the amplifier backaction may prove to be the Achilles' heel. In this work, we develop an ab-initio theoretical framework to understand the high frequency dynamics of the SQUID in detail. In addition to giving us crucial insights into the amplifying mechanism of the MWSA and its nonreciprocal response between the input and output signal channels, this approach enables us to calculate the experimentally relevant quantities such as available gain, added noise and directionality at operating frequencies of interest.
\par 
We perform our analysis in the paradigm of input-output theory and employ the method of harmonic balance to study the driven dynamics of the device. The dc SQUID is biased in the voltage regime --- in contrast to the usual Josephson parametric amplifiers operated in the zero voltage state with the phase excursions of the Josephson junction confined to a single cosine well --- and has the dynamics of a particle sampling various wells of a two-dimensional tilted washboard \cite{SQUIDvol1}. The input-output analysis thus needs to be generalized to take into account phase running evolution in this two-dimensional potential. Our approach involves a self-consistent determination of the working point of the device established by static bias parameters (the static bias current $I_{B}$ and external flux $\Phi_{\rm ext}$ shown in Fig. \ref{Fig1}) followed by a study of the rf dynamics using a perturbative series expansion around this working point. In Sec. \ref{sec_model} we introduce our input-output model for the dc SQUID. Using this in Sec. \ref{sec_acdc}, we first derive the response at zero frequency and at the Josephson oscillation frequency $\omega_{J}$ in a self consistent manner. Following this, we evaluate the perturbative response at finite frequency around zero and $\omega_{J}$ as a scattering matrix in the basis of relevant modes of the circuit, which clearly elucidates the nonreciprocal dynamics of the device. In Secs. \ref{sec_Gp} and \ref{sec_TN} we calculate the figures of merit such as power gain, directionality and noise temperature, and identify the fundamental limits on the performance of the device. Section \ref{sec_concl} contains our concluding remarks.
%
\section{Analytical Model}
\label{sec_model}
%
\subsection{SQUID circuit basics}
%
\begin{figure}
\centering
  \includegraphics[width=0.65\columnwidth]{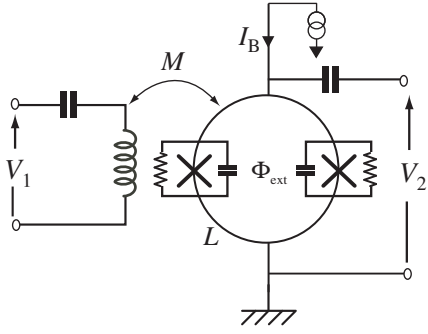}\\
  \caption{Circuit schematic of a conventional MWSA. The SQUID consists of two Josephson junctions, arranged in a superconducting loop, with inductance $L$. The loop is biased using a static current source $I_{B}$ and an external flux $\Phi_{\rm ext}$. An input voltage $V_{1}$ generates an oscillating current in an input coil inductively coupled to the SQUID thus inducing a small flux modulation $\delta \Phi$ of the flux enclosed by the loop. For optimal flux bias [$\Phi_{\rm ext} = (2n+1)\Phi_{0}/4$] that maximizes the flux-to-voltage transfer coefficient $V_{\Phi} \equiv (\partial V/\partial \Phi_{\rm ext})_{I_{B}}$, this causes an output voltage $V_{2}= V_{\Phi}\delta \Phi$ to develop across the ring. Thus the device behaves as a low impedance voltage amplifier.}\label{Fig1}
\end{figure}
\begin{figure*}
  \includegraphics[width=0.7\textwidth]{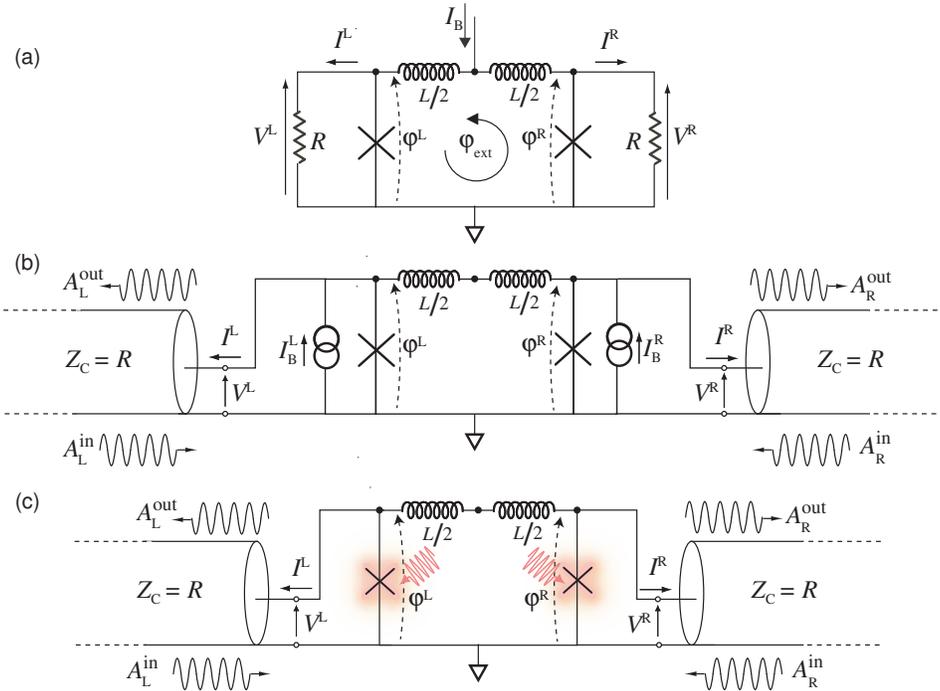}\\
  \caption{Equivalent input-output model of the SQUID (a) The bare SQUID loop without the input coupling circuit. As noted in Fig. \ref{Fig1}, the circuit has two static biases - a common mode bias current $I_{B}$ and a differential mode external flux $\Phi_{\rm ext}$. There is also a capacitance $C$ across each junction, not shown here for simplicity. (b) Equivalent SQUID circuit under Nyquist representation of shunt resistances and separate static current biases $I_{B}^{L}$ and $I_{B}^{R}$ for the left and right junction respectively. The common mode bias current $I_{B}$ now corresponds to the even combination of two external bias currents $-(I_{B}^{L} + I_{B}^{R})/2$ while the external flux $\Phi_{\rm ext}$ corresponds to the differential combination of the two current sources $L(I_{B}^{L} - I_{B}^{R})/2$. The oscillating signals are modelled as incoming and outgoing waves travelling on semi-infinite transmission lines, representing the shunt resistances across the two junctions. (c) Effective junction representation for evaluating the signal response of the device. Here, we have replaced the junctions biased with a static current with effective junctions pumped by the Josephson harmonics (represented by a ``glowing" cross with a pumping wave) generated by phase running evolution in the voltage state of the junction.} \label{Fig2}
\end{figure*}

The SQUID circuit considered in our analysis is shown in Fig. \ref{Fig2}(a). The dynamics of the system are modelled as a particle moving in a two-dimensional potential of the form \cite{SQUIDvol1}
\begin{eqnarray}
    \frac{U_{\rm SQUID}}{2 E_{J}}(\varphi^{D}, \varphi^{C}) &=& \frac{1}{\pi \beta_{L}} \left(\varphi^{D} - \frac{\varphi_{\rm ext}}{2}\right)^{2}
    \nonumber\\
    & & \quad - \; \cos\varphi^{D}\cos\varphi^{C} - \frac{I_{B}}{2 I_{0}} \varphi^{C}.\;\;
    \label{squidpot}
\end{eqnarray}
Here, $I_{B}$ is the bias current, $I_{0}$ is the critical current of each junction, $\varphi_{\rm ext}= 2 \pi \Phi_{\rm ext}/\Phi_{0}$ represents the externally imposed flux in the loop, $\beta_{L} \equiv 2 L I_{0}/\Phi_{0}$ denotes a dimensionless parametrization of the SQUID loop inductance, $E_{J} \equiv I_{0}\Phi_{0}/2 \pi$ is the Josephson energy and $\Phi_{0} \equiv h/2e$ is the flux quantum. We have introduced the common, $\varphi^{C} = (\varphi_{L} + \varphi_{R})/2$, and differential, $\varphi^{D} = (\varphi_{L} - \varphi_{R})/2$, mode combinations of the phases of the two junctions that form the axes of the two-dimensional orthogonal coordinate system.
\par
To facilitate an input-output analysis of the circuit, we replace the resistive shunts across the junction with semi-infinite transmission lines [cf. Fig. \ref{Fig2}(b)] of characteristic impedance $Z_{C}=R$, following the Nyquist model of dissipation. Thus the shunts play the dual role of dissipation and ports (or channels) used to address the device. This allows us to switch from a standing mode representation in terms of lumped element quantities such as voltages and currents to a propagating wave description in terms of signal waves travelling on the transmission lines. The amplitude of these waves is given by the well known input-output relation \cite{IOT},
\begin{equation}
    A_{i}^{\rm in/out}(t) =\frac{V^{i}}{2\sqrt{Z_{C}}} \mp \frac{\sqrt{Z_{C}} I^{i}}{2}, \quad [i \in \{L, R\}].
    \label{IOT}
\end{equation}
where $V^{i}$ and $I^{i}$ denote the voltage across the shunt resistance and current flowing in the shunt resistance respectively.
From Eq. (\ref{squidpot}), we obtain the common mode current, $I^{C} = (I^{L} + I^{R})/2$, and differential mode current, $I^{D} = (I^{L} - I^{R})/2$, flowing in the shunts by identifying $\varphi^{C,D}$ as the relevant position variables. The current in each mode can thus be interpreted as the ``force" \cite{MichelQMckts} that follows directly from Hamilton's equation of motion as
\begin{eqnarray}
    \frac{I^{C,D}}{I_{0}} = \frac{\partial}{\partial \varphi^{C,D}}\left(\frac{U_{\rm SQUID}}{E_{J}}\right),
\end{eqnarray}
which yields
\begin{subequations}
\begin{align}
    \widehat{\omega}^{C} & = \frac{\omega_{B}}{2} - \omega_{0} \sin\varphi^{C}\cos\varphi^{D}\footnotemark\\
    \widehat{\omega}^{D} & = \frac{\omega_{0}}{\pi \beta_{L}}(-2 \varphi^{D} + \varphi_{\rm ext}) 
                    - \omega_{0} \cos\varphi^{C}\sin\varphi^{D}.
\end{align}
    \label{currents}
\end{subequations}
Here we have expressed the currents in equivalent frequency units,
\begin{eqnarray}
    \widehat{\omega}^{C} \equiv \frac{I^{C} R}{\varphi_{0}} \; & {\rm and} & \;
    \widehat{\omega}^{D} \equiv \frac{I^{D} R}{\varphi_{0}}, \quad (\rm currents)\\
    \omega_{B} \equiv \frac{I_{B} R}{\varphi_{0}} \; & {\rm and} & \; \omega_{0} \equiv \frac{I_{0} R}{\varphi_{0}},
        \quad (\rm characteristic\;\;currents) \nonumber\\
\end{eqnarray}
with $\varphi_{0} = \Phi_{0}/(2\pi)$. Including a capacitance across the junction gives an additional term, involving a second-order derivative of the common and differential mode fluxes, of the form $-\omega_{B} \Omega_{c} \ddot{\varphi}^{C,D}$ with $\Omega_{c} = 2 \pi I_{B} R^{2} C/\Phi_{0}$, on the right-hand side of Eq. (\ref{currents}). This parametrization of capacitance, motivated by calculational simplicity, leads to a different parametrization of the plasma frequency, $\omega_{p} \equiv (I_{0}/\varphi_{0} C)^{1/2}[1 - (I_{B}/I_{0})^{2}]^{1/4}$. The more conventional parametrization with a fixed value of capacitance for all bias values can be implemented in a more comprehensive calculation aided by numerical techniques. 
%
%
\par
Equation (\ref{currents}) represents a subtle current-phase relationship for the two-junction system, analogous to the first Josephson relation. We note that Eqs. (\ref{currents}) can alternatively be derived using a first-principles Kirchoff law analysis of the circuit in Fig. \ref{Fig2}(a). Similar to the currents, we can define the common and differential mode voltages as
\begin{equation}
    \widecheck{\omega}^{C} \equiv \frac{V^{C}}{\varphi_{0}}; \quad \widecheck{\omega}^{D} \equiv \frac{V^{D}}{\varphi_{0}}.
    \label{voltages}
\end{equation}
Further, by the second Josephson relation, we have
\begin{eqnarray}
    \langle \widecheck{\omega}^{C}\rangle =  V_{dc}/\varphi_{0} = \omega_{J},
    \label{vdcwj}
\end{eqnarray}
where $V_{dc}$ is the static voltage developed across the SQUID biased in the running state.
\par
We note that the usual mode of operation of a dc SQUID involves an input flux inductively coupled using an input transformer of which the loop inductance forms the secondary coil (Fig. \ref{Fig1}). The input transformer, however, is an experimental artifact required to ensure the impedance matching with the input impedance of the SQUID at a desired frequency. It is not crucial from the point of view of device characteristics, however, as it is the SQUID which provides amplification and all the relevant nonlinear dynamics of the device. In the ensuing analysis we do not employ a separate input port, but rather consider a direct input coupling through the differential mode of the ring which couples to the flux in an analogous manner [Fig. \ref{Fig2}(b)]. Such a scheme may also prove beneficial for a practical device to overcome the problem of low coupling at high signal frequencies, as recently shown experimentally using a SLUG (Superconducting Low-inductance Undulatory Galvanometer) microwave amplifier \cite{ribeill:103901,hover:063503}. 
%
\subsection{Harmonic balance treatment}
\begin{figure}
  \includegraphics[width=\columnwidth]{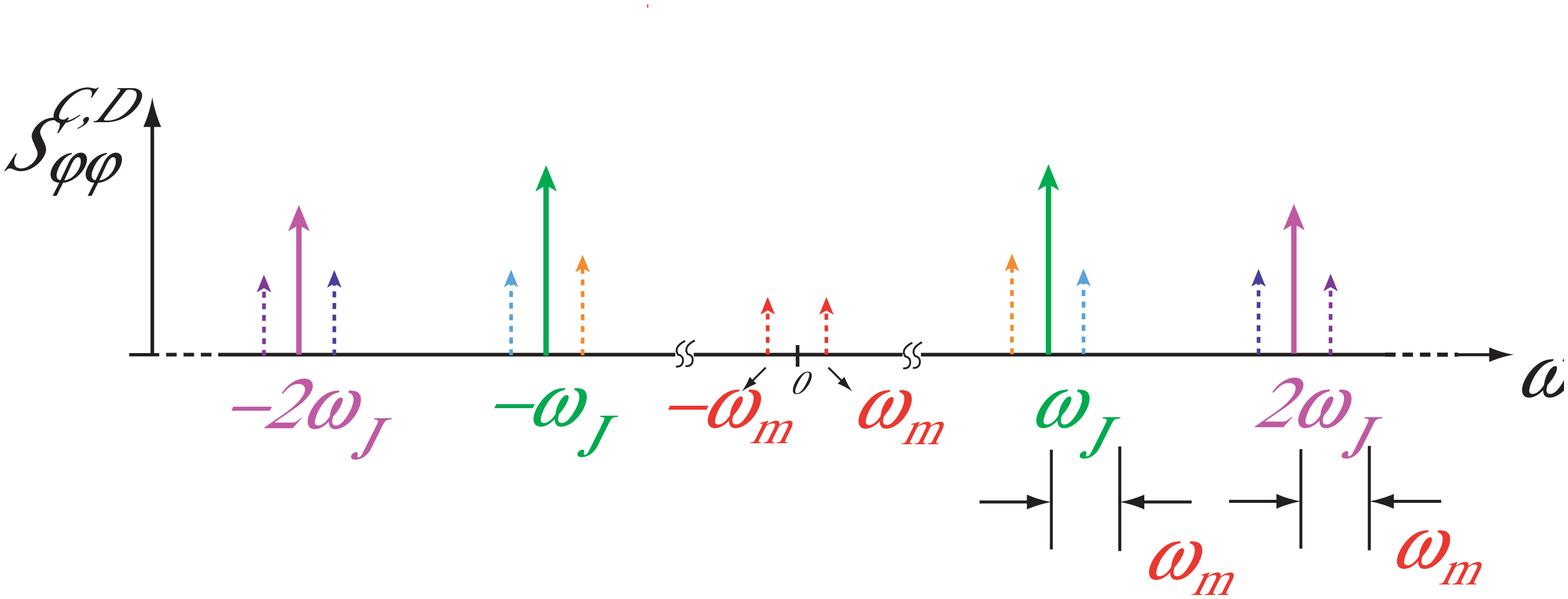}\\
  \caption{(Color online) Spectral density landscape of common and differential modes of the SQUID. The tall solid arrows show the Josephson harmonics generated internally in the running state of the device. The small input signal frequency $\omega_{m}$ and different sidebands generated by mixing with Josephson harmonics are shown with dashed arrows.}\label{FigSpec}
\end{figure}

%
%
Using the input-output relation of Eq. (\ref{IOT}) with Eqs. (\ref{currents}) and (\ref{voltages}), we obtain the equations 
\begin{equation}
    \widecheck{\omega}^{C,D}(t) = \widehat{\omega}^{C,D}(t) + 2 \omega^{{\rm in} C,D}(t)
    \label{EOM}
\end{equation}
for common and differential mode circuit quantities. We employ the technique of harmonic balance and solve Eq. (\ref{EOM}) in the frequency domain, at all frequencies of interest (see Fig. \ref{FigSpec}). This is achieved by assuming two parts to the solution for each variable of interest ($\varphi^{C}$ and $\varphi^{D}$),
\begin{eqnarray}
    \varphi^{C} &=& \omega_{J} t + \delta\varphi^{C}(t) 
    \label{phiE}\\
    \varphi^{D} &=& \phi_{0} + \delta\varphi^{D}(t),
    \label{phiD}
\end{eqnarray}
where $\omega_{J}t$ and $\phi_{0}$ represent the average static values of the common and differential mode phases [cf. Eq. (\ref{vdcwj})]. The time varying components are of the form
\begin{eqnarray}
    \delta\varphi^{C,D}(t) = \Pi^{C,D}(t) + \Sigma^{C,D}(t),
\end{eqnarray}
where $\Pi(t)$ refers to the components at the Josephson frequency $\omega_{J}$ and its harmonics. The term $\Sigma(t)$ includes the components oscillating at the signal frequency $\omega_{m}$ and its resultant sidebands $\omega_{n} = n \omega_{J} + \omega_{m}$ generated by wave mixing via the nonlinearity of the SQUID:
\begin{eqnarray}
    \Pi^{C,D} &=& \sum_{k=1}^{K} p_{k,x}^{C,D} \cos k\omega_{J} t + p_{k,y}^{C,D}\sin k\omega_{J} t \label{pumpsandsignals1}\\
    \nonumber\\
    \Sigma^{C,D} &=& \sum_{n=-N}^{+N} s_{n,x}^{C,D}\cos(n\omega_{J}+ \omega_{m})t \nonumber\\
    & & \qquad \; \; + s_{n,y}^{C,D}\sin(n\omega_{J} + \omega_{m})t.\;\;
    \label{pumpsandsignals2}
\end{eqnarray}
\par
We note that the number of Josephson harmonics included in the analysis [i.e. $K$ in Eq. (\ref{pumpsandsignals1})] is determined by the order of expansion of the junction nonlinearity in $\delta\varphi$. This in turn is determined by the bias voltage of the device set by the bias current $I_{B}$. As $I_{B}$ is reduced towards the critical current of the junction $I_{0}$, higher Josephson harmonics become more significant as the characteristics of the device become increasingly nonlinear. We can, therefore, calculate the response perturbatively by expanding each of the coefficients $p$ and $s$ in Eqs. (\ref{pumpsandsignals1})-(\ref{pumpsandsignals2}) as a truncated power series in the reduced bias parameter  
\begin{equation}
    \varepsilon \equiv \frac{I_{0}}{I_{B}} = \frac{\omega_{0}}{\omega_{B}}.
\end{equation}
The degree of the resultant polynomial evaluation of $p, \; s$ coefficients is set by the desired order of expansion in $\delta\varphi$. As $\varepsilon \leq 0.5$ (or equivalently $I_{B} > 2 I_{0}$) for the SQUID to operate in the running state at any value of flux bias \cite{SQUIDvol1}, which is the regime of interest for the SQUID to be operated as a voltage amplifier, it provides a convenient small parameter of choice ensuring rapid convergence of the perturbation series method. Furthermore, this parameter serves as the effective strength of the different Josephson harmonics which play a role analogous to the strong ``pump" tone of conventional parametric amplifiers.
%
%
%
\section{Calculation of SQUID Dynamics}
\label{sec_acdc}
%
%
\subsection{Steady state response: I -- V characteristics}
\label{subsec_dc}
%
%
\begin{figure}[t!]
  \includegraphics[width=0.95\columnwidth]{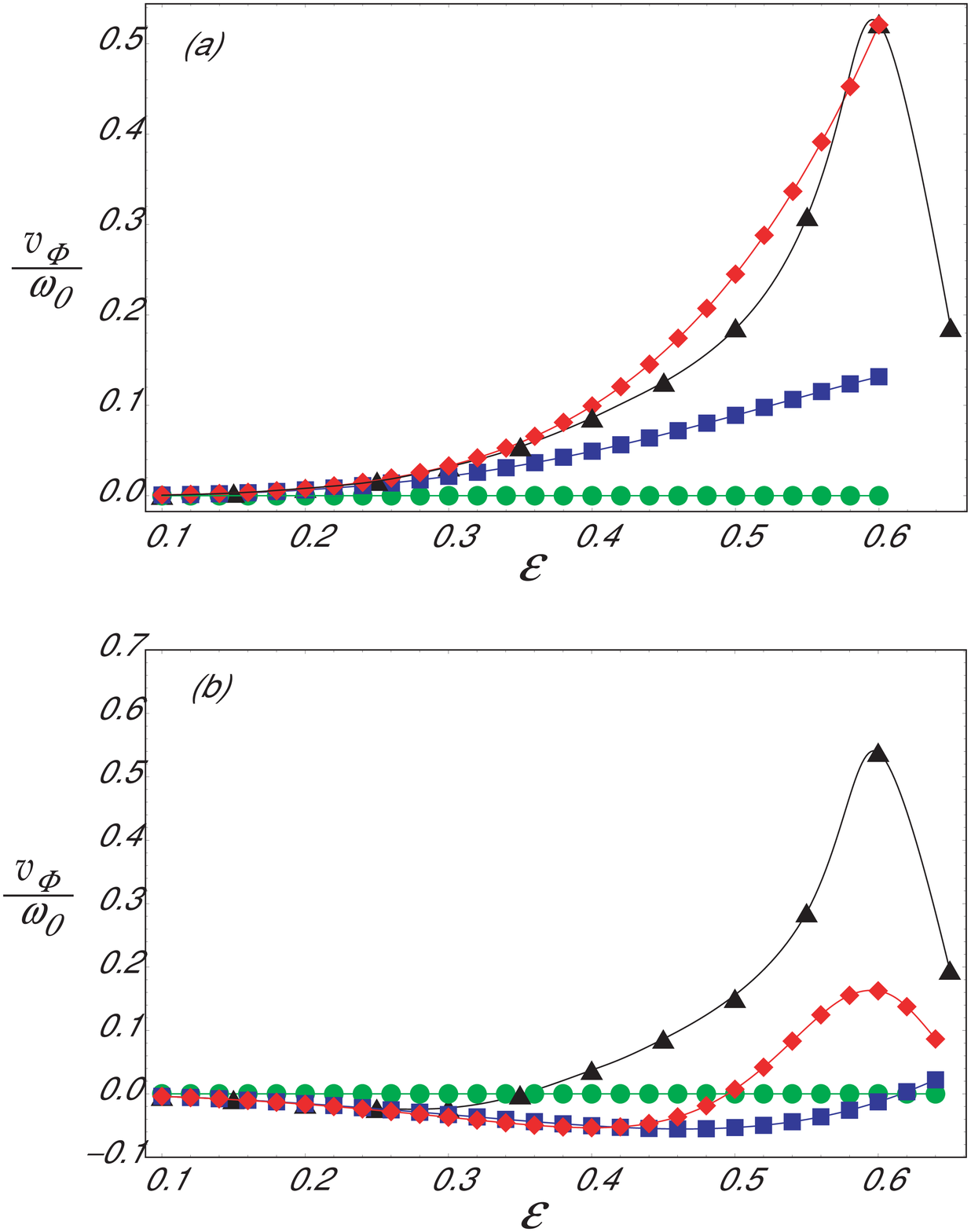}\\
  \caption{(Color online) Static transfer function of the SQUID calculated as a function of two bias parameters $\varepsilon = \omega_{0}/\omega_{B}$ at $\varphi_{\rm ext} = \pi/2$ for (a) strongly overdamped ($\Omega_{C} =0$) and (b) intermediately damped junctions ($\Omega_{C} =1$). The (black) triangles represent the transfer function calculated from the exact numerical integration of the SQUID equations. The (green) circles correspond to the $K=1$ evaluation including only the Josephson frequency [Eq. (\ref{pumpsandsignals1})]. This first order evaluation does not show any voltage modulation with flux as there is no coupling between the common and differential modes at this order. The (blue) squares and (red) diamonds correspond to an evaluation including the second ($K=2$) and third Josephson ($K=3$) harmonic respectively. The corresponding curves represent interpolating polynomials. In both plots, the agreement of the perturbative series with the exact numerical solution improves on including higher order corrections corresponding to contributions of higher Josephson harmonics.}\label{Fig3}
\end{figure}
We first determine the working point of the SQUID by solving for the steady state characteristics. As the zero frequency response of the system is intrinsically related to the response at the Josephson frequency through Eq. (\ref{vdcwj}), we calculate it self-consistently along with the strength of the various Josephson harmonics in the steady state by considering only the static source terms with no oscillating input drive at the Josephson frequency and its harmonics. This yields a set of boundary conditions of the form
\begin{eqnarray}
    & & \widecheck{\omega}[k \omega_{J}] - \widehat{\omega}[k \omega_{J}] = 0 \quad k \in [0,K].  
\end{eqnarray}
We solve this set of simultaneous equations to calculate the strength of the various Josephson harmonics generated internally from the static bias due to the junction nonlinearity along with the zero frequency characteristics. Figure \ref{Fig3} shows a plot of the static transfer function $\upsilon_{\Phi} = \partial \langle\omega^{C}\rangle / \partial \varphi_{ext}$ obtained using the perturbative series method to determine the coefficients $p_{k}^{C,D}$ [Eq. (\ref{pumpsandsignals1})] described in the last section. The agreement between the exact numerical calculation and the perturbative analytical calculation improves on increasing the order of the perturbation series expansion by including mixing processes mediated by higher Josephson harmonics. Further, from the steady state calculation for the differential mode, we obtain a relation for the phase angle between the two junctions in the ring as
\begin{eqnarray}
    \phi_{0} = \frac{\varphi_{\rm ext}}{2} + \beta_{L}\sum_{k \geq 2}^{K} a_{k} \varepsilon^{k} \sin\varphi_{\rm ext},
    \label{phi0}
\end{eqnarray}
where the coefficients $a_{k}$ are of order unity. Thus, we see that the average values of both the explicit static bias parameters namely $\varepsilon$ (common) and $\varphi_{\rm ext}$ (differential) participate in establishing each of the implicit static biases -- $V_{dc}$ (or equivalently $\omega_{J}$) for the common mode and $\phi_{0}$ for the differential mode. The contributions arising from the bias current, as shown in Eq. (\ref{phi0}), lead to a rolling of the static phase difference around the SQUID loop that manifests itself as the change in curvature of the transfer function curves shown in Fig. (\ref{Fig3}). Furthermore, we note that, as indicated by the steady state calculation, the flux dynamics of $\upsilon_{\Phi}$ evaluated using the truncated harmonic series calculation are `slower', that is, they shift to higher values of bias with respect to the exact numerical results. Nonetheless, the predicted magnitudes are comparable and hence the theory is capable of making semi-quantitative predictions in an analytically tractable manner. The major merit of this approach over conventional methods lies in the natural extension offered for the study of higher frequency dynamics as discussed in the following sections.
%
%
\subsection{RF response: Scattering Matrix}
%
%
Once we have determined the static working point for the SQUID, we can solve for its rf dynamics in the small signal regime. The aim is to calculate signal amplitudes by including the $\Sigma^{C,D}(t)$ term in our analysis and considering all the mixing processes mediated by the pumps $\Pi^{C,D}(t)$ evaluated in the last section, permissible by the harmonic balance of Eqs. (\ref{currents}), (\ref{voltages}) and (\ref{EOM}). This is equivalent to the representation shown in Fig. \ref{Fig2}(c), where we model the mixing of the input signal by the SQUID as a parametric interaction with different Josephson harmonics playing the role of an effective ``colored" pump. In the limit of a small amplitude input signal, which is the relevant limit for most practical situations, we can then introduce a linear response description of the dynamics as an admittance matrix seen from the ports. This can be obtained from the current-phase and voltage-phase relationship [Eqs. (\ref{currents}),(\ref{voltages})] as
\begin{eqnarray}
    \overrightarrow{\widecheck{\omega}} = \widecheck{\mathbb{M}}\overrightarrow{\Sigma}
    \label{Vtophi}
\end{eqnarray}
and
\begin{eqnarray}
    \overrightarrow{\widehat{\omega}} = \widehat{\mathbb{M}}\overrightarrow{\Sigma},
    \label{Itophi}
\end{eqnarray}
yielding
\begin{eqnarray}
    \mathbb{Y} = \widehat{\mathbb{M}}\widecheck{\mathbb{M}}^{-1}.
    \label{Ymat}
\end{eqnarray}
The vectors in the equations above are defined in the basis of all signal and sideband frequencies of interest, $(\Sigma^{C}[n\omega_{J}+\omega_{m}], \Sigma^{D}[n\omega_{J}+\omega_{m}]),\; \; n \in [-N, +N]$ leading to a $4(2N+1) \times 4(2N+1)$ admittance matrix. We further note that the matrix is block diagonal since harmonic balance leads to two disjoint manifolds, each of which forms a closed subspace of dimension $2(2N+1)$.
\begin{figure*}[t!]
  \includegraphics[width=0.95\textwidth]{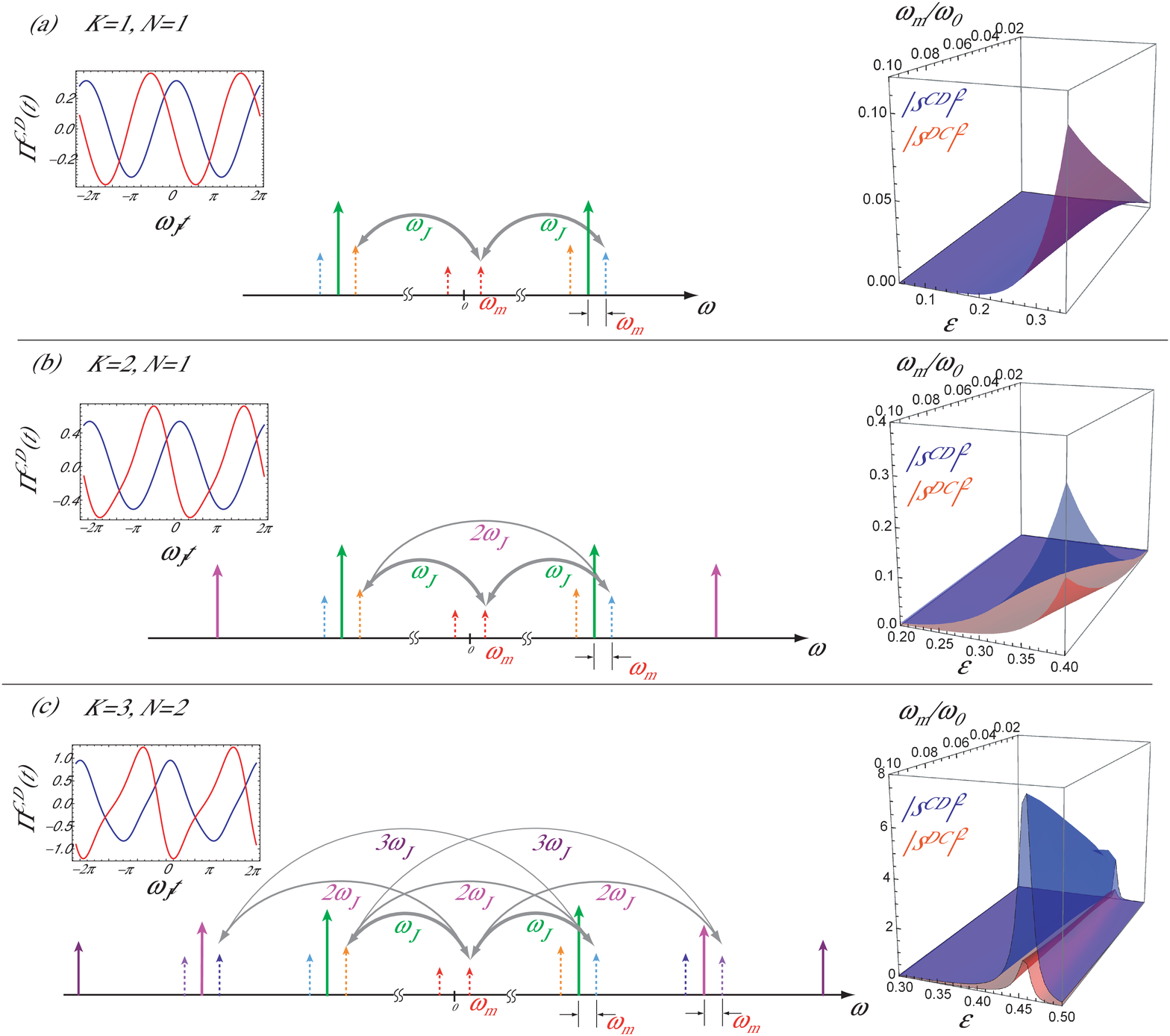}\\
  \caption{Josephson harmonics and small signal scattering gain of the SQUID calculated using harmonic balance with expansion of the $\sin\varphi$ nonlinearity to (a) first, (b) third and (c) fifth order respectively. Each panel shows the relevant modes of the frequency spectrum included in the calculation at that order [Eqs. (\ref{pumpsandsignals1})-(\ref{pumpsandsignals2})]. The dispersive mixing between various temporal modes of the system is denoted using grey arcs with the relevant Josephson harmonic acting as the pump indicated next to them. The relative strength of the different mixing processes is indicated by the respective widths of the arcs, with the strongest being denoted by the thickest arcs. Also shown are plots of $\Pi(t)$, the effective pumps in common (blue) and differential (red) modes at each order of the calculation. The box panels show the respective forward ($|s^{CD}|^{2}$) and backward ($|s^{DC}|^{2}$) scattering gains as a function of reduced input frequency $\omega_{m}/\omega_{0}$ and bias parameter $\varepsilon = \omega_{0}/\omega_{B}$. The surface plot in (a), calculated using only the Josephson frequency, shows no asymmetry between the forward and backward gains (blue and red surface plots respectively). The asymmetry develops on inclusion of higher harmonics that implement a multitone pump which is not symmetric about $t=0$, as seen from the plots of $\Pi(t)$ in panels (b) and (c). As we increase the order of calculation and include higher harmonics, the asymmetry increases and finally peaks at an optimal value of bias parameter $\varepsilon = 0.455$.}\label{Figscatt}
\end{figure*}
\par
From the admittance matrix of Eq. (\ref{Ymat}), we can evaluate the scattering matrix of the SQUID using the identity
\begin{eqnarray}
    \mathbb{S} = (\mathbb{U} + \mathbb{Y})^{-1}(\mathbb{U} - \mathbb{Y}).
    \label{scattmat}
\end{eqnarray}
Figure \ref{Figscatt} shows the calculation for different orders in junction nonlinearity and the relevant forward ($|s^{CD}|^{2}$) and backward scattering gain ($|s^{DC}|^{2}$). It immediately shows the emergence of the nonreciprocal gain of the device that, unlike conventional paramps, enables a two-port operation. As the nonlinearity of the device characteristics is increased by reducing $I_{B}$ towards $I_{0}$ (thus increasing the expansion parameter $\varepsilon$), we need to include the higher Josephson harmonics in the calculation which become significant due to rapid running evolution of the phase of the junctions in the two-dimensional tilted washboard. This leads to a situation analogous to pumping of the SQUID by an effective multitone pump of the form $\Pi(t) = \sum_{k=1}^{K} p_{k} \cos(\omega_{J}t+\phi_{k})$ in both $C$ and $D$ modes [see $\Pi(t)$ panels in Fig. \ref{Figscatt}]. The dynamics of such a system include multi-path interference involving different Josephson harmonics. This effect, analogous to symmetry breaking in ratchet physics \cite{Hanggi1}, implements an asymmetric frequency conversion scheme guided by relative phases $\phi_{k}$ of  different Josephson harmonics driving the junctions \cite{ArchanaRSJ}. The signal in the differential mode is preferentially upconverted, coupled through higher order mixing processes into the common mode and then preferentially downconverted into the common mode, yielding a net forward gain  from the differential mode to the common mode. The reverse gain process from $C$ to $D$ is disfavored by the same reasoning, leading to the nonreciprocal operation of the SQUID amplifier.
%
%
\section{Power Gain of the SQUID}
\label{sec_Gp}
%
The dc SQUID operated as a two-port voltage amplifier resembles the configuration of a semiconductor, operational amplifier (op-amp) as opposed to that of a conventional parametric amplifier, which is a matched device. (That is, the input and output impedances are identical to the impedances of the transmission lines or coaxial cables). In this sense the MWSA is the magnetic dual of the rf SET (single electron transistor) \cite{DevoretSET}. The dc SQUID amplifies an input current (directly coupled as in this analysis or coupled as a flux via an input transformer), and has a much lower impedance than the electromagnetic environment in which it is embedded. Conversely, the rf SET amplifies an input voltage, and has a much higher impedance than the electromagnetic environment in which it is embedded. The true power gain of either device, as seen from the ports, thus involves a de-embedding of the device characteristics. In the case of the SQUID, this requires a translation from the matched (or scattering) description based on the input-output theory considered in this paper to the op-amp or hybrid representation that is well suited for describing an unmatched amplifier such as the microwave MWSA.
\par
The hybrid matrix describing a two-port amplifier is of the form \cite{ClerkRMP}
\begin{eqnarray}
    \left(\begin{array}{c}
    V_{2} \\
    I_{1}
    \end{array}\right)
    =
    \left(
    \begin{array}{cc}
    \lambda_{V} & Z_{\rm out}\\
    Y_{\rm in} & \lambda_{I}^{'}
    \end{array}
    \right)
    \left(\begin{array}{c}
    V_{1} \\
    I_{2}
    \end{array}\right).
    \label{2portamp}
\end{eqnarray}
where $(V_{1}, I_{1})$ and $(V_{2}, I_{2})$ denote the voltage and current associated with the input and output ports respectively. The power gain for such an amplifier is given by
\begin{eqnarray}
    G_{P} = \frac{P^{\rm out}}{P^{\rm in}} &=& \frac{V_{2}^{2}/{\rm Re}[Z_{\rm out}]}{V_{1}^{2}/{\rm Re}[Z_{\rm in}]}\nonumber\\
    &=& \frac{\lambda_{V}^{2}}{{\rm Re}[Y_{\rm in}]{\rm Re}[Z_{\rm out}]},
    \label{GPeq}
\end{eqnarray}
where $\lambda_{V}$ is the voltage gain of the amplifier, $Y_{\rm in}$ is the input admittance and $Z_{\rm out}$ is the output impedance.  Equation (\ref{GPeq}) represents the gain of an effective ``matched" device accounting for the impedance mismatch at the input and output ports. 
%
\par
In principle, although the calculation of quantities in Eq. (\ref{2portamp}) can be performed using the scattering matrix evaluated in Eq. (\ref{scattmat}) \cite{ClerkRMP}, nonetheless it is advantageous to transform to a description that is more natural in describing the relationship between standing mode current and voltage variables. We find that an impedance matrix ($\mathbb{Z}$) representation is well suited for such a purpose due to its rather straightforward mapping to the standing mode quantities of Eq. (\ref{2portamp}). Using the $\mathbb{Y}$-matrix, derived in Eq. (\ref{Ymat}), we can write the impedance matrix $\mathbb{Z}$ of the dc SQUID as
\begin{eqnarray}
    \mathbb{Z} &=& (\mathbb{U} + \mathbb{Y})^{-1}
    \label{EqZmat}
\end{eqnarray}
with
\begin{eqnarray}
    \left(\begin{array}{c}
    \overrightarrow{\widecheck{\omega}^{C}} \\
    \overrightarrow{\widecheck{\omega}^{D}}
    \end{array}\right)
    =
    \left(
    \begin{array}{cc}
    z^{CC} & z^{CD}\\
    z^{DC} & z^{DD}
    \end{array}
    \right)
    \left(\begin{array}{c}
    \overrightarrow{\widehat{\omega}^{C}}\\
    \overrightarrow{\widehat{\omega}^{D}}
    \end{array}\right).
    \label{Zmat}
\end{eqnarray}
Here, as before, $\overrightarrow{\widecheck{\omega}}$, $\overrightarrow{\widehat{\omega}}$ are vectors defined in the space of all signal and sideband frequencies of interest. Also $\mathbb{U}$ is an identity matrix of appropriate dimensions and corresponds to the admittance contribution of the resistive shunts across the junctions. 
\par
\begin{figure}
  \includegraphics[width=0.95\columnwidth]{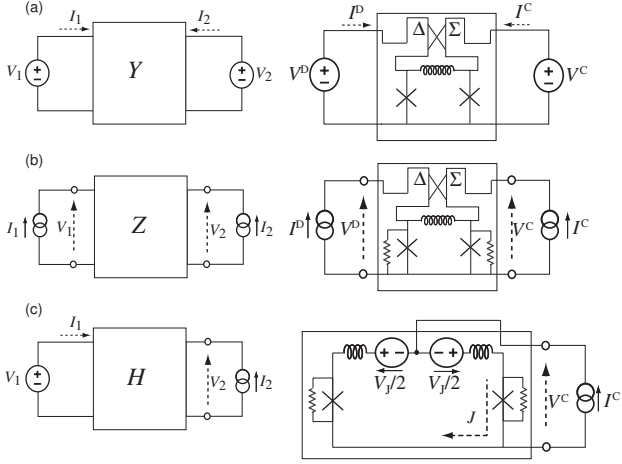}\\
  \caption{Different representations of a two-port network and analog configurations for the dc SQUID. (a) $\mathbb{Y}$-matrix representation defined for closed boundary conditions $Y_{ij} = dI_{i}/dV_{j}|_{V_{k \neq j}=0}$ for the junctions and inductance, omitting the shunts [Eq. (\ref{Ymat})]. (b) $\mathbb{Z}$-matrix representation defined for open boundary conditions $Z_{ij} = dV_{i}/dI_{j}|_{I_{k \neq j}=0}$ including the shunts [Eq. (\ref{EqZmat})]. (c) (Hybrid) $\mathbb{H}$-matrix or op-amp representation defined with mixed boundary conditions [Eq. (\ref{2portamp})]. In effective matrices for the SQUID, the common mode (C) and differential mode (D) excitations of the ring play the role of ports 1 and 2, if the SQUID is addressed using hybrids. In each panel, the quantities shown with solid arrows represent the stimulus while those shown with dashed arrows represent the corresponding response of the network.}\label{Fig2portreps}
\end{figure}

The next step is to make the translation from the impedance matrix derived in the common and differential mode basis to the two-port description of Eq. (\ref{2portamp}). This requires an identification of the correct``input" and ``output" voltages and currents for the circuit in Fig. \ref{Fig2}(a). As the SQUID readout involves measurement of the voltage developed across it, the relevant output quantities are related to the common mode quantities as $V_{2} = V^{C}$ and $I_{2} = 2 I^{C}$. The translation to the input variables of the hybrid representation is more subtle. For this purpose we first note that, in conventional SQUID operation, the input flux coupled into the ring modulates the circulating current $J$ which is, thus, the relevant input current of the device. The equivalent input voltage that causes the flux modulation of the circulating current can be represented by a voltage source $V_{J}$ in series with the inductance of the loop. Figure \ref{Fig2portreps} summarizes the different possible two-port representations of the SQUID used in this paper.
\par
\begin{figure}[t!]
\centering
\includegraphics[width=0.9\columnwidth]{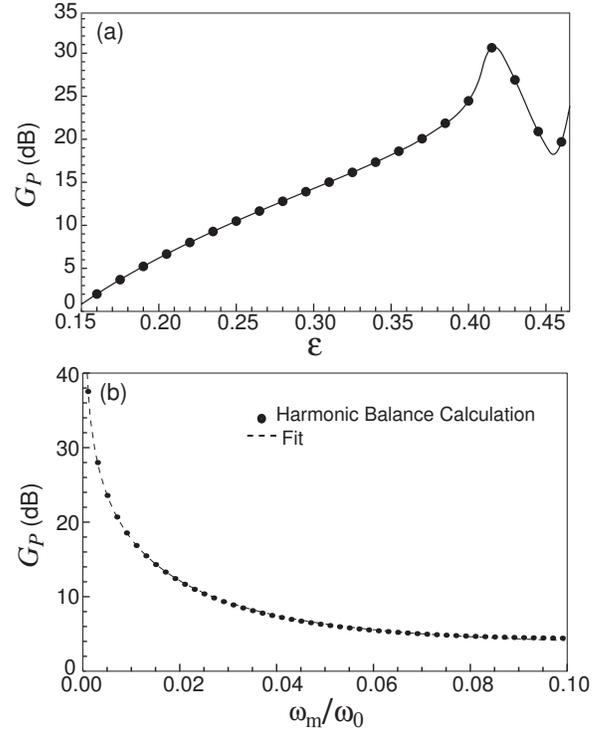}\\
\caption{Power gain of the MWSA calculated with $K=3$, $N=2$, taking into account the modification of the input and output impedances of the device by matched loads. The parameters are $\Phi=\Phi_{0}/4$, $\beta_{L} = 1$ and $\Omega_{C} = 1$. (a) Power gain versus bias parameter $\varepsilon = \omega_{0}/\omega_{B}$ calculated for a fixed input frequency $\omega_{m} = 0.01 \;\omega_{0}$. The solid curve is an interpolating polynomial of degree two. (b) Power gain versus input frequency $\omega_{m}/\omega_{0}$ calculated with bias parameter fixed at $\varepsilon = 0.455$, the optimum value for attaining minimum noise temperature [see Fig. \ref{FigTN}(a)] at low frequencies ($\omega \ll \omega_{0}$).  The fit is of the form $G_{P} = [0.006/(\omega_{m}/\omega_{0})^{2}] + 2$ measured in linear units.}
\label{FigGp}
\end{figure}

On interpreting the loop variables $(V_{J},J)$ described above in terms of the differential mode voltage $V^{D}$ and current $I^{D}$ (see appendix \ref{app_hyb} for details), we obtain the following equivalence between the coefficients of the hybrid matrix in Eq. (\ref{2portamp}) and the $\mathbb{Z}$-matrix of Eq. (\ref{Zmat}):
\begin{eqnarray}
    \lambda_{V} &=& \left(\frac{R}{i\omega_{m} L}\right)z^{CD}[\omega_{m}]\\
    \lambda_{I} &=& \left(\frac{R}{i\omega_{m} L}\right)z^{DC}[\omega_{m}]\\
    Z_{\rm out} &=& \left(\frac{R}{2}\right) z^{CC}[\omega_{m}]\\
    Y_{\rm in} &=& (i\omega_{m} L)^{-1} + \left(\frac{2R}{\omega^{2} L^{2}}\right) z^{DD}[\omega_{m}].
\end{eqnarray}
Using the above translation in Eq. (\ref{GPeq}), we find an expression for the power gain purely in terms of $\mathbb{Z}$-matrix coefficients:
\begin{eqnarray}
    G_{P}[\omega_{m}] = \frac{|z^{CD}[\omega_{m}]|^{2}}{{\rm Re}[z^{CC}[\omega_{m}] {\rm Re}[z^{DD}[\omega_{m}]]}.
    \label{GpfromZ}
\end{eqnarray}
Figure \ref{FigGp} shows the power gain of the device as a function of bias and input frequency, calculated using Eq. (\ref{GpfromZ}). It shows that power gain of the MWSA increases quadratically with decreasing input signal frequency, a result corroborated by a simple quasistatic treatment presented in Appendix \ref{app_dcequi}. The reverse power gain of the device is calculated in a similar manner as 
\begin{eqnarray}
    G_{P}^{\rm rev} &=& \frac{I_{1}^{2}{\rm Re}[Z_{\rm in}]}{I_{2}^{2}{\rm Re}[Z_{\rm out}]}
    = \frac{\lambda_{I}^{2}}{{\rm Re}[Y_{\rm in}]{\rm Re}[Z_{\rm out}]}\nonumber \quad [{\rm Eq.} \;(\ref{2portamp})]\\
    &=& \frac{|z^{DC}[\omega_{m}]|^{2}}{{\rm Re}[z^{CC}[\omega_{m}] {\rm Re}[z^{DD}[\omega_{m}]]}.
\end{eqnarray}
\begin{figure}[t!]
  \includegraphics[width=0.95\columnwidth]{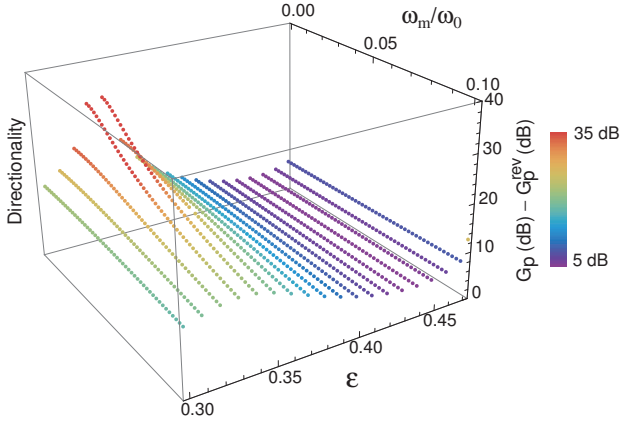}\\
  \caption{Directionality (in dB) of the MWSA as a function of bias parameter $\varepsilon = \omega_{0}/\omega_{B}$ and reduced input frequency $\omega_{m}/\omega_{0}$. The red dots represent high directionality and blue dots represent low directionality. The parameters used were the same as those in Fig. \ref{FigGp}.}\label{Figisolation}
\end{figure}
\par
The directionality ($G_{P} - G_{P}^{\rm rev}$) --- which is a measure of the asymmetry between forward and reverse power gains --- follows directly from the asymmetric scattering gain discussed in the previous section. Our calculation shows that it is a strong function of the bias $\varepsilon$ (Fig. \ref{Figisolation}); furthermore the optimal bias for maximum power gain is not the same as that for maximal directionality. We note that the results presented here have been obtained with a truncated harmonic series excluding all Josephson harmonics above $3 \omega_{J}$. In the real device, the achievable isolation between forward (differential-to-common) and backward (common-to-differential) gain channels may be quantitatively different due to the presence of the neglected higher order interferences.
%
%
\section{Noise Temperature}
\label{sec_TN}
%
%
In this section, we evaluate the noise added by the dc SQUID operated as a voltage amplifier. The noise added by a system can be quantified by its noise temperature, $T_{N}$, defined as
\begin{eqnarray}
    T_{N} = A\frac{\hbar \omega}{k_{B}}
\end{eqnarray}
where $A$ is the Caves added noise number \cite{Caves}. This noise temperature corresponds to the effective input temperature of the amplifier obtained by referring the added noise measured at the output to the input, and is quantified in terms of energy quanta per photon at the signal frequency. For a phase preserving amplifier, such as the MWSA, the minimum possible noise temperature corresponds to half a photon of added noise, that is, $A_{\rm min}=0.5$ in the large gain limit (in general, $A_{\rm min}= 1/2 -1/2G$). 
\par
Using the hybrid representation developed in the previous section and Appendix \ref{app_hyb}, we write the noise inequality for the MWSA as
\begin{eqnarray}
    k_{B}T_{N} \geq \frac{\sqrt{\bar{S}_{V^{C}V^{C}}\bar{S}_{JJ} - {\rm Re}[\bar{S}_{V^{C}J}]^{2}}- {\rm Im}[\bar{S}_{V^{C}J}]}{\lambda_{V}},
    \label{TNdef}
\end{eqnarray}
where $\bar{S}_{VV}$ represents the spectral density of the voltage fluctuations at the output, $\bar{S}_{JJ}$ represents the spectral density of the circulating current fluctuations and $\bar{S}_{VJ}$ is the cross-correlation between the voltage and current fluctuations \cite{ClerkRMP}. 
\par
As in the case of power gain, we can evaluate these quantities from the $\mathbb{Z}$ matrix of the SQUID derived in Sec. \ref{sec_Gp}. This exercise is enabled by the fact that the input-output theory treats the deterministic signal input and noise of the system on an equal footing. Thus, the linear response description developed to calculate the signal gain provides a straightforward way of generalizing the theory to understand the noise properties of the system, simply by replacing the input current signal with a noise signal described by a spectral density of the form
\begin{eqnarray}
    \bar{S}_{II} [\omega] = 2\hbar\omega {\rm Re}[Y[\omega]]\coth\left(\frac{\hbar\omega}{2k_{B}T}\right).
    \label{noisespec}
\end{eqnarray}
As before, we use the $\mathbb{Z}$ matrix to calculate the spectral densities in Eq. (\ref{TNdef}). Using this we write the voltage noise spectral density in the common mode as
\begin{eqnarray}
    \bar{S}_{V^{C}V^{C}} &=& \sum_{n=-N}^{N}|z_{0n}^{CC}|^{2} \bar{S}_{I^{C}I^{C}}[n\omega_{J}+\omega_{m}] \nonumber\\
    & & \;\;\; + \sum_{n=-N}^{N}|z_{0n}^{CD}|^{2} \bar{S}_{I^{D}I^{D}}[n\omega_{J}+\omega_{m}].
    \label{SVV}
\end{eqnarray}
Here, the first sum accounts for the contribution to the noise arising from the common mode signal ($n=0$) and sidebands about the Josephson harmonics ($n=\pm 1, \pm 2$) included in the calculation; the second sum accounts for the noise generated in the common mode output signal by the differential mode signal and sidebands, arising from coupling between $C$ and $D$ modes. Similarly, we can calculate $\bar{S}_{JJ}$ as
\begin{eqnarray}
    \bar{S}_{JJ} = \frac{4}{\omega_{m}^{2} L^{2}} \bar{S}_{V^{D}V^{D}},
\end{eqnarray}
by making the identification $J = 2 V^{D}/(i \omega L)$. Here, as before, we calculate $S_{V^{D}V^{D}}$ from the $Z$ matrix:
\begin{eqnarray}
    \bar{S}_{V^{D}V^{D}} &=& \sum_{n=-N}^{N}|z_{0n}^{DC}|^{2} \bar{S}_{I^{C}I^{C}}[n\omega_{J}+\omega_{m}] \nonumber\\
    & & \;\;\; + \sum_{n=-N}^{N}|z_{0n}^{DD}|^{2} \bar{S}_{I^{D}I^{D}}[n\omega_{J}+\omega_{m}].
    \label{SDD}
\end{eqnarray}
Finally, for $S_{VJ}$ we have
\begin{eqnarray}
 \bar{S}_{V^{C}J} & = & \frac{-2}{i \omega_{m} L}\left(\sum_{n=-N}^{N} z_{0n}^{CC}z_{0n}^{DC*} \bar{S}_{I^{C}I^{C}}[n\omega_{J}+\omega_{m}] \right.   \nonumber\\
 & & \;\; \left.+ \sum_{n=-N}^{N} z_{0n}^{CD}z_{0n}^{DD*} \bar{S}_{I^{D}I^{D}}[n\omega_{J}+\omega_{m}]\right).
 \label{SVD}
\end{eqnarray}
\begin{figure}[t!]
  \includegraphics[width=0.9\columnwidth]{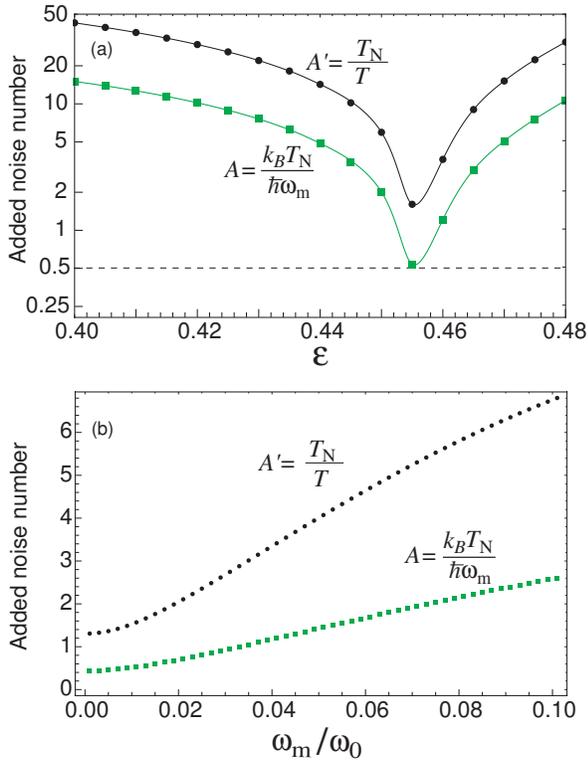}\\
  \caption{(Color online) Caves added noise number for the MWSA, calculated using the harmonic balance analysis with $K=3$, $N=2$ as a function of (a) $\varepsilon = \omega_{0}/\omega_{B}$ for $\omega_{m} = 0.01 \; \omega_{0}$ and (b) reduced input frequency $\omega_{m}/\omega_{0}$ with $\varepsilon = 0.455$. In both plots, (round) black markers show the noise number $A^{'} = T_{N}/T$ obtained in the thermal regime $k_{B}T \gg \hbar \omega_{m}$, while (square) green markers show the noise number $A=k_{B} T_{N}/\hbar \omega_{m}$ calculated in the quantum regime $k_{B} T \ll \hbar \omega_{m}$. The solid curves in (a) represent interpolating polynomials. The quantum calculation gives a minimum value for $A = k_{B} T_{N}/ \hbar \omega_{m} \approx 0.5$, attained at $\varepsilon = 0.455$, corresponding to one-half photon of added noise [horizontal black dashed line in (a)]. The optimal value of bias current for minimum added noise does not coincide with that for achieving the maximum power gain [Fig. \ref{FigGp}(a)] or directionality (Fig. \ref{Figisolation}).}\label{FigTN}
\end{figure}
\par
Figure \ref{FigTN} shows plots of the Caves noise number of the device, calculated in both the thermal regime [$k_{B}T \gg \hbar \omega_{m}$, where all terms in Eqs. (\ref{SVV})-(\ref{SVD}) contribute equal noise powers] and the quantum regime [$k_{B}T \ll \hbar \omega_{m}$, where each term in Eqs. (\ref{SVV})-(\ref{SVD}) contributes a noise power proportional to its frequency in accordance with Eq. (\ref{noisespec})]. 
\par
There are a number of points to be highlighted. Our calculation shows that, at the optimal current bias of $\varepsilon = 0.455$, the MWSA attains the quantum limit of added noise corresponding to half-photon at signal frequency. Moreover, the optimum bias point for minimum noise corresponds to the bias for maximum scattering gain [Fig. \ref{Figscatt}(c)] rather than for the maximum power gain [Fig. \ref{FigGp}(a)]. This result, previously found both theoretically and experimentally \cite{JohnLesHouches}, follows from the fact that the added noise is a property of the bare SQUID without any matching to input and output loads. In the case of conventional parametric amplifiers, the minimum noise indeed occurs at the maximum scattering gain. Furthermore, the partial cross-correlation between the output voltage noise across the SQUID and the supercurrent noise circulating in the loop is crucial to minimizing the noise in both thermal and quantum regimes. We also note that, for sufficiently low signal frequencies, the calculated added noise number is found to saturate at a value slightly below the quantum limit of one half-photon at the signal frequency. This result, we suspect, is due to the fact that at the bias for minimum noise, the reverse gain is substantial and hence the isolation is not perfect (Fig. \ref{Figisolation}). The quantum limit of one half-photon is a limiting value calculated for ideal detectors with zero reverse gain and high forward gain \cite{ClerkRMP}, a condition which is not satisfied at the optimal noise bias in our calculation. Finally, our calculation shows that the minimum noise number is achieved only when the signal frequency is much lower than the characteristic Josephson frequency $\omega_{0} = 2 \pi I_{0} R/\Phi_{0}$, and increases significantly with increasing signal frequency.
%
%
\section{Concluding remarks}
\label{sec_concl}
%
In summary, we have developed a new method based on input-output theory to provide a first-principles analysis of the microwave SQUID amplifier (MWSA). In this paradigm we treat the SQUID biased in its running state as a parametric amplifier pumped by a combination of Josephson harmonics generated internally by the motion of the phase of the junctions. This approach leads to a fully self-consistent description of both the static and rf dynamics of the device. The scattering matrix calculation shows that the nonreciprocal gain of the amplifier arises from mixing processes involving higher Josephson harmonics which implement an asymmetric frequency conversion scheme involving upconversion to and subsequent downconversion from the Josephson frequency. We find that the power gain of the matched SQUID amplifier decreases quadratically with signal frequency $\omega_{m}$; by comparison, the gain in the usual SQUID operation with a matched input coil scales as $1/\omega_{m}$ \cite{JohnLesHouches}. However, a recently reported dc SQUID amplifier \cite{hover:063503} using the direct coupling method considered in this paper demonstrated that the power gain scaled as $1/\omega_{m}^{2}$ at a frequency of a few GHz and a bandwidth of several hundred MHz.
\par
Our analysis shows that the MWSA achieves quantum-limited noise performance for optimal flux and current biases, and for signal frequencies significantly lower than the characteristic Josephson frequency $\omega_{0} = 2 \pi I_{0}R/\Phi_{0}$. The added noise increases significantly with increasing frequency. This problem can be alleviated by using junctions with higher values of critical currents. With the present technology for niobium junctions, critical current densities of tens of microamperes per square micron are readily achievable. This translates into characteristic frequencies of about 100 GHz, which should be sufficient to achieve lower noise at GHz frequencies provided hot electron effects due to dissipation in the shunts are mitigated \cite{PhysRevB.49.5942}. Furthermore, our analysis shows that simultaneous optimization of gain, directionality and noise is a delicate operation since the optimal biases for these three properties do not coincide. Based on our calculation, at the working point for minimum added noise, $A \thickapprox 0.5$, power gains of $15-18$ dB and directionality of around $5-8$ dB are obtained. However, higher power gains of $20-30$ dB and directionality of $10-12$ dB can be realized by permitting a higher noise number $A \thickapprox 5-10$. Though the predicted directionality is still modest, it suffices to reduce the number of nonreciprocal elements (circulators, isolators) in the measurement chain typically employed for the readout of superconducting qubits. Moreover, the noise penalty incurred with MWSAs compares very well to standard cryogenic amplifiers such as HEMTs whose typical noise numbers lie in the range $40-50$ for microwave frequencies.
\par
Although the results presented in this paper are semi-quantitative we believe that extension of the analysis to higher orders, in conjunction with numerical optimization techniques, can be a useful tool to analyze SQUID-based devices due to rapid convergence offered by the harmonic series method. This approach would allow one to evaluate the appropriate parameters, depending on the intended application, that yield the best compromise between gain and noise properties.
%
%
\begin{acknowledgements}
\label{sec_acknow}
We thank O. Buisson, A. Clerk, M. Hatridge and R. Vijay for useful discussions. The authors also wish to thank Ananda Roy for help with the calculations in initial stages of this project. This research was supported by ARO under Grant No. W911NF-09-1-0514 (AK and MHD) and the U.S. Department of Energy, Office of High Energy Physics, under contract number DE-FG02-11ER41765 (JC). 
\end{acknowledgements}
%
%
%
\appendix 
%
\section{Loop Variables for the dc SQUID}
\label{app_hyb}
%
%
In this appendix we establish the correspondence between the differential mode variables that serve as the input in the analysis of Secs. \ref{sec_model} and \ref{sec_acdc} and the input variables required for the hybrid representation discussed in Sec. \ref{sec_Gp}. The output variables in the two representations have a simple relationship as explained in Sec. \ref{sec_Gp}. 
\begin{figure}[h!]
  \includegraphics[width=0.75\columnwidth]{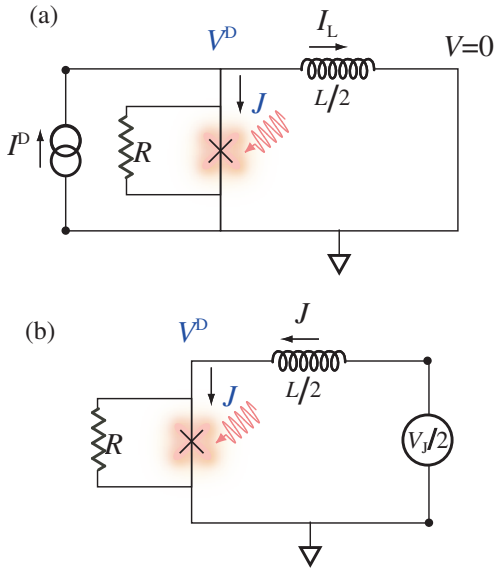}\\
  \caption{(Color online) Equivalence between the SQUID differential mode and the op-amp input variables. Here, for simplicity, we have used the symmetric version of the SQUID to divide the ring along the equipotential. The circuit in (a) models the device as an impedance response function to an imposed current source $I^{D}$, as a result of which a voltage drop $+V^{D} (-V^{D})$ develops across the left (right) junction. (b) Hybrid representation for the SQUID, which models the input response by introducing a differential voltage source in the SQUID loop and recording the current that flows through the junction.  The junction at this point is replaced with an effective junction pumped using the various Josephson harmonics generated by the static bias current [also see Fig. \ref{Fig2}(c)].}\label{FigHybckt}
\end{figure}
\par
Figure \ref{FigHybckt} shows the two representations [Figs. \ref{Fig2portreps}(b) and (c)], one in terms of differential mode quantities ($V^{D}, I^{D}$) suitable for a scattering or matched representation (since the input and output impedances are just the transmission line impedance) and the other in terms of a circulating current $J$ and a loop voltage $V_{J}$, which are the relevant input quantities for the device in an unmatched hybrid description. In Fig. \ref{FigHybckt}(a) Kirchoff's current law gives
\begin{eqnarray}
    J = I^{D} - I_{L},
    \label{Ypic}
\end{eqnarray}
while in Fig. \ref{FigHybckt}(b), from Kirchoff's voltage law, we have
\begin{eqnarray}
    -\frac{V_{J}}{2} = V^{D} + \frac{i \omega L}{2} J,
    \label{Hypic}
\end{eqnarray}
with $J = I_{L}$. 
\par
To establish the equivalence of the two representations from the point of view of the junction, we require the voltage across the junction $V^{D}$ and current through the junction $J$ to be conserved (see Fig. \ref{FigHybckt}). Thus, using Eq. (\ref{Ypic}) in Eq. (\ref{Hypic}), we obtain
\begin{eqnarray}
    V_{J} = -i \omega L I^{D}.
\end{eqnarray}
Similarly it is easily seen that the circulating current $J$ is given as
\begin{eqnarray}
    J = \frac{2 V^{D}}{i \omega L}.
\end{eqnarray}
%
%
\section{Static analog circuit for the SQUID}
\label{app_dcequi}
\begin{figure}
  \includegraphics[width=0.7\columnwidth]{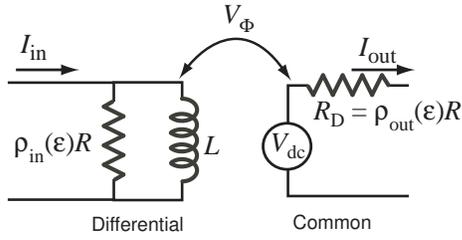}\\
  \caption{Equivalent low frequency circuit for a SQUID for calculation of unilateral power gain. The input circuit is modelled as an effective impedance viewed by a low frequency differential mode current. The output circuit impedance comprises a bias-dependent resistor, denoting the dynamic impedance of the junction, that converts the output voltage to a corresponding output current. The net ``transimpedance" is given by the static flux-to-voltage transfer function of the device. The symbols $\rho_{\rm in, \; out}$ denote bias-dependent constants of order unity.}\label{FigEquickt}
\end{figure}

\begin{figure}[h!]
  \includegraphics[width=0.9\columnwidth]{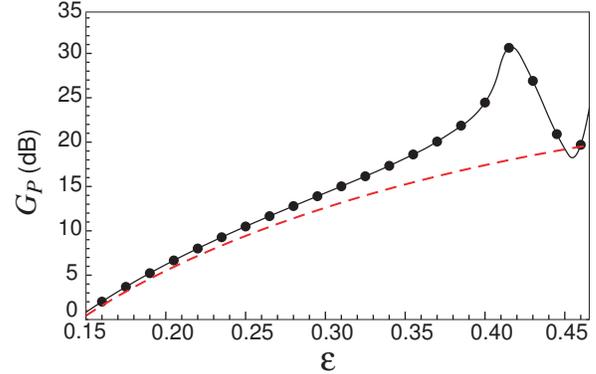}\\
  \caption{Comparison of the power gain as a function of bias $\varepsilon = \omega_{0}/\omega_{B}$ calculated using a $K=3, \; N=2$ calculation (solid black line) and a purely quasistatic calculation (dashed red line) of Eq. (\ref{Gpdcfinal}). For the quasistatic gain calculation, $V_{\Phi}$ and $\rho_{\rm in,out}$ were obtained from the $I-V$ characteristics evaluated in Sec. \ref{subsec_dc}.}\label{Figacdcequi}
\end{figure}

%
%
The SQUID can be thought of as a current amplifier with a current transferred from a low-impedance input port to a high-impedance output port. This description is analogous to the FET dual model with the gain given by a transimpedance instead of a transconductance. The equivalent `current gain' of such a device (Fig. \ref{FigEquickt}) for frequencies sufficiently close to zero [$\omega_{m} \ll \rho_{\rm in}R/L = \omega_{0}/(\pi \beta_{L})$ to be precise] can be modelled as
\begin{eqnarray}
        \frac{I_{\rm out}}{I_{\rm in}} \approx \frac{V_{\Phi}L}{R_{D}}.
\end{eqnarray}
This leads to a power gain
\begin{eqnarray}
        G_{P}^{{\rm dc}} = \left(\frac{I_{\rm out}}{I_{\rm in}}\right)^{2} \frac{R_{D}}{{\rm Re}[Z_{\rm in}]}.
        \label{Gpdc}
\end{eqnarray}
For frequencies of interest, ${\rm Re}[Z_{\rm in}] \approx \omega_{m}^{2} L^{2}/(\rho_{\rm in}R)$. Using this result in Eq. (\ref{Gpdc}), we obtain the power gain
\begin{eqnarray}
        G_{P}^{{\rm dc}} = \frac{\rho_{\rm in}}{\rho_{\rm out}}\left(\frac{V_{\Phi}}{\omega_{m}}\right)^{2},
        \label{Gpdcfinal}
\end{eqnarray}
which can be rewritten as,
\begin{eqnarray}
    G_{P}^{{\rm dc}} \approx \rho_{g} \left(\frac{\omega_{0}}{\omega_{m}}\right)^{2},
    \label{finalGp}
\end{eqnarray}
where $\rho_{g}$ is a bias-dependent and frequency-independent constant of order unity. Here, we have used the relation $V_{\Phi}^{\rm opt} = R/L = \omega_{0}/\pi$ \cite{SQUIDvol2} for $\beta_{L}=1$. Equation (\ref{finalGp}) shows that the gain drops quadratically with increasing signal frequency, and that no power gain is obtained for signal frequencies close to the plasma frequency of each junction in the SQUID. This frequency dependence of the power gain is borne out by the full rf analysis shown in Fig. \ref{FigGp}(b).  Figure \ref{Figacdcequi} shows a comparison of gain calculated using quasistatic response functions as shown in Eq. (\ref{Gpdcfinal}) and a rf calculation at low frequencies, involving the third Josephson harmonic [same as that shown in Fig \ref{FigGp}(a)]. The agreement is better for lower values of $\varepsilon$ where high frequency components of the device are less significant. The impedance matrix calculation generates extra terms due to self-summation caused by the inversion operation [cf. Eq. (\ref{EqZmat})] which leads to higher order corrections absent from the purely quasistatic calculation.
%
%
%

%
%
\end{document}